\documentclass[12pt]{article}
\pdfoutput=1

\usepackage{amssymb}
\usepackage{amsmath}
\usepackage{amscd}
\usepackage{mathrsfs}
\usepackage{amssymb}
\usepackage{feynmf}
\usepackage{graphicx}
\usepackage[pdftex,bookmarks=true,pdfauthor={},pdftitle={},pagebackref=false]{hyperref}
\usepackage{youngtab}
\usepackage{color}
\usepackage{mdframed}
\usepackage{breakurl}
\usepackage{cite}

\numberwithin{equation}{section}

\begin{document}

\setlength{\oddsidemargin}{0cm}
\setlength{\baselineskip}{7mm}

\thispagestyle{empty}
\setcounter{page}{0}

\begin{flushright}

\end{flushright}

\vspace*{-1cm}

\begin{center}
{\bf \Large }
\end{center}

\begin{center}
{\bf \Large Sudakov Representation of the Cachazo-He-Yuan} 

\vspace*{0.3cm}

{\bf \Large Scattering Equations Formalism }

\vspace*{0.2cm}

\vspace*{1cm}

Grigorios Chachamis$^{a,}$\footnote{\tt  chachamis@gmail.com}, Diego Medrano Jim\'enez$^{a,}$\footnote{\tt  d.medrano@csic.es},  
Agust\'{\i}n Sabio Vera$^{a,}$\footnote{\tt 
a.sabio.vera@gmail.com} \\
and Miguel \'A. V\'azquez-Mozo$^{b,}$\footnote{\tt 
Miguel.Vazquez-Mozo@cern.ch}

\end{center}

\vspace*{0.0cm}

\begin{center}

$^{a}${\sl Instituto de F\'{\i}sica Te\'orica UAM/CSIC \&
Universidad Aut\'onoma de Madrid\\
C/ Nicol\'as Cabrera 15, E-28049 Madrid, Spain
}

$^{b}${\sl Departamento de F\'{\i}sica Fundamental, 
 Universidad de Salamanca \\ 
 Plaza de la Merced s/n,
 E-37008 Salamanca, Spain
  }
\end{center}

\vspace*{1.5cm}

\centerline{\bf \large Abstract}
We show that the use of Sudakov variables greatly simplifies the study of the solutions to the scattering equations 
in the Cachazo-He-Yuan formalism. We work in the center-of-mass frame for the two incoming particles, which partially
fixes the ${\rm SL}(2,{\mathbb C})$ redundancy in the integrand defining the scattering amplitudes, the remaining freedom translates into a global shift in the azimuthal angle of the outgoing particles. Studying four- and five-particle amplitudes, we see how an appropriate choice of this phase allows for algebraic simplifications when finding solutions to the scattering equations, as well as in the expression of the scattering amplitudes in terms of the locations of the punctures in the sphere. These punctures themselves are remarkably simple functions of the Sudakov parameters. 

\noindent

\newpage  

\setcounter{footnote}{0}

\section{Introduction}

Scattering amplitudes are the building blocks for the computation of observables in 
quantum field theory and string theory (see \cite{Elvang:2015rqa} for a review). The progress in their understanding and calculation in 
recent years has been enormous. Usual Feynman diagram techniques become too complicated as the number of external legs or loops 
increases and any alternative procedures are always desirable. The Cachazo-He-Yuan (CHY) formalism~\cite{Cachazo:2013hca,Cachazo:2013iea}  
represents a very promising route in this direction. It presents $n$-point amplitudes as an $(n-3)$-dimensional integral over the moduli 
space of $n$-punctured spheres, fully localized on the solutions to the so-called {\sl scattering equations} 
(SE)~\cite{Cachazo:2013hca,Gross:1987kza,Gross:1987ar,Fairlie:2008dg,Fairlie:1972zz}. A proof of this formula for 
arbitrary $n$ was given by Dolan and Goddard \cite{Dolan:2013isa}. The CHY proposal for the calculation of tree-level scattering 
amplitudes has an interpretation in terms of ambitwistor 
strings~\cite{Mason:2013sva,Berkovits:2013xba,Adamo:2013tsa,Ohmori:2015sha,Casali:2015vta} defined on a Riemann surface at genus zero. At 
loop level, supergravity integrands of four-point amplitudes at one and two loops have been obtained when introducing higher 
genus~\cite{Adamo:2013tsa,Adamo:2015hoa,Geyer:2015bja,Geyer:2015jch}. Other connections to string theory amplitudes can be found 
in, for example,~\cite{Bjerrum-Bohr:2014qwa,Gomez:2013wza,Casali:2014hfa}. 

The main difficulty with the CHY strategy is that the number of integrals defining the $n$-point $S$-matrix elements, although 
being compensated by a delta function, grows very rapidly. The reason is that the number of integrations to be carried out 
goes like the number of 
solutions to the SE, which is $(n-3)!$ for the $n$-point amplitude. There has been steady progress in 
the understanding of the solutions to the SE and the calculation of 
amplitudes obtained from them (see, for example, 
\cite{Weinzierl:2014vwa,Dolan:2014ega,Lam:2014tga,Kalousios:2015fya,Baadsgaard:2015voa,Baadsgaard:2015ifa,Cardona:2015ouc,Dolan:2015iln}). 

In this work we focus on the physical interpretation of the solutions to the SE in terms of the positions of the associated 
punctures on the Riemann sphere. We find that Sudakov variables~\cite{Sudakov:1954sw}, which parametrize outgoing momenta in terms of its projections on two incoming momenta 
and a vector transverse to their collision axis, are a very efficient 
way to present the solutions to the SE, since they naturally encode momentum conservation. When evaluating the scattering amplitudes it is 
also useful to work in the center-of-mass frame of the two incoming particles. This is equivalent to partially fixing the 
SL(2,${\mathbb C}$) redundancy, localizing two of the punctures at opposite poles of the sphere while
leaving a third puncture free. This residual symmetry corresponds to the freedom in the choice of the origin for the azimuthal angle with respect to the axis defined by the incoming particles. 
Choosing this global phase wisely allows for a simple representation of the scattering amplitude in terms of the position of the 
punctures on the sphere, which also admit a simple representation. 

In Section 2, we review those aspects of the CHY approach which are of special interest for our work. In particular, we 
discuss 
in detail the solution to the SE found in \cite{Fairlie:2008dg}, which exists for any number of external particles in four 
dimensions and which we write in terms of the rapidities and the azimuthal angles of the emitted particles. In Section 3, 
after identifying two of the particles participating in the scattering as incoming, we work in their center-of-mass frame 
taking the $z$ axis as their direction of flight. This is done
through a double scaling limit involving the rapidities and transverse momenta. 

Section 4 is devoted to describe the use of Sudakov variables in the simple case of four-particle scattering. 
The punctures associated with the outgoing momenta
are characterized by a single Sudakov variable and one azimuthal angle, which parametrizes circles on the Riemann sphere. 
Then we 
calculate the four-point amplitude for a scalar cubic theory using this representation. 
In Section 5 we analyze the more complicated case of the 
five-point amplitudes. In this case four Sudakov variables and two azimuthal angles are needed to parametrize the system of SE and the punctures positions. We 
show how to obtain a second solution to the SE as the complex conjugate of the one previously discussed. At the end of this section we 
evaluate the corresponding amplitude for a scalar theory and express it as a simple function of the Sudakov variables. Finally, 
in Section 6 we present our conclusions and directions for future work. 

\section{The CHY formalism for scattering amplitudes}

\subsection{Momentum space and the punctured sphere}

In a scattering problem the data are codified in a set of 
$n$ on-shell $D$-dimensional momenta $p_i^{\mu}$ satisfying energy-momentum conservation 
\begin{align}
\sum_{i=1}^n p_i=0,
\end{align}
modulo Lorentz transformations.
The departing point of the CHY formalism \cite{Cachazo:2013hca,Cachazo:2013iea} is a mapping from these momenta 
into an internal space on the $n$-punctured Riemann sphere parametrized by the variables 
$\sigma_i \in \mathbb{C}\mathbb{P}^1$ (with $i=1,\ldots,n$) through the identity 
\begin{align}
p^\mu_j  = \oint\limits_{\left|z-\sigma_j \right|=\epsilon} \frac{dz}{2 \pi i} \frac{v^\mu(z)}{\prod_{k=1}^n (z-\sigma_k)},
\end{align}
where 
\begin{align}
v^\mu(z) = \sum_{j=1}^{n}p_j^\mu \prod_{\substack{k=1\\k\neq j}}^n (z-\sigma_k).
\end{align} 
Due to momentum conservation, this is a vector-valued polynomial of degree $n-2$ satisfying $v(z)^{2}=0$. 
This latter null condition for $v(z)$ implies the SE 
\begin{align}
{\cal S}_i (\sigma) \equiv \sum_{j \neq i}^n \frac{s_{ij}}{\sigma_{ij}}=0,
\label{eq:S_i=0}
\end{align}
where $\sigma_{ij} \equiv \sigma_i - \sigma_j$ and we have introduced the Mandelstam invariants
\begin{align}
s_{ij}=(p_{i}+p_{j})^{2}=2p_{i}\cdot p_{j}. 
\label{eq:mandelstam_inv}
\end{align} 
Although there are $n$ equations, only $n-3$ are linearly independent 
as a consequence of SL(2,$\mathbb{C}$) invariance. These SE have a total of $(n-3)!$ solutions mapping the space of 
kinematic invariants into $(n-3)!$ points in the moduli space of $n$-punctured spheres. 

The SE first appeared in Ref. \cite{Fairlie:1972zz} in the study of the ground state configuration for the Koba-Nielsen representation of scattering amplitudes of open strings,
\begin{align}
A_n = \int  d\sigma_2 \dots d\sigma_{n-2} \prod_{\substack{i,j=1\\i>j }}^{n-1}  \sigma_{i j}^{-2\alpha' p_i \cdot p_j},
\end{align}
where $0=\sigma_1 < \sigma_2 < \dots < \sigma_{n-1}=1$. The dominant saddle-point region was investigated by Gross and Mende~\cite{Gross:1987kza,Gross:1987ar} in the closed string case and by Gross and Ma\~nes for open strings~\cite{Gross:1989ge}.
In both cases, all $s_{ij}$ are taken to be large simultaneously, corresponding precisely to Eq. \eqref{eq:S_i=0}.

\subsection{Scattering amplitudes}

Remarkably, as shown in \cite{Cachazo:2013hca,Cachazo:2013iea}, all tree-level $n$-point Yang-Mills amplitudes in $D$ dimensions can be obtained from the following integral representation with support on the solutions to the SE: 
\begin{align}
{\cal A}_n = i \, g^{n-2} \int \frac{d^{n} \sigma}{{\rm Vol} [{\rm SL}(2,\mathbb{C})]}
\sigma_{kl} \sigma_{lm} \sigma_{mk} {\prod_{i \neq k,l,m}}
\delta \left(\sum_{j \neq i}^n \frac{2 \, p_i \cdot p_j}{\sigma_{ij}}\right) I_L I_R.
\label{eq:general_amplitude_SE}
\end{align}
The invariance of this expression under SL(2,$\mathbb{C}$) transformations of the $\sigma_{i}$ guarantees that the result is independent of the
choice of $\{k,l,m\}$. 
In the integrand, $I_{L}$ carries the color traces 
\begin{align}
I_L &= \sum_{\beta \in S_n/\mathbb{Z}_n} 
\frac{
{\rm Tr} 
\left(
T^{a_{\beta(1)}} T^{a_{\beta(2)}} \cdots T^{a_{\beta(n)}}
\right)
}{
\sigma_{\beta(1) \beta(2)} \sigma_{\beta(2) \beta(3)} \cdots \sigma_{\beta(n) \beta(1)}
},
\label{eq:IL}
\end{align}
with the sum running over non-cyclic permutations. The second factor
\begin{align}
I_R ~= {\rm Pf}' M_n,
\label{eq:IR_redpfaffian}
\end{align}
is the reduced Pfaffian
of the $2n\times 2n$ antisymmetric matrix
\begin{align}
M_n &= \left(\begin{array}{cc}
M_A & - M_C^T\\
M_C & M_B
\end{array}\right).
\end{align}
where the block matrices $M_{A}$, $M_{B}$, and $M_{C}$ are given by
\begin{align}
M_{A}^{ij} &= \left\{\begin{array}{cll}
\frac{p_i \cdot p_j}{\sigma_{ij}} & {\rm for} & i \neq j\\[0.2cm]
0 & {\rm for} & i = j
\end{array} \right. , \nonumber \\[0.2cm]
M_{B}^{ij} &= \left\{\begin{array}{cll}
\frac{\epsilon_i \cdot \epsilon_j}{\sigma_{ij}} & {\rm for} & i \neq j\\[0.2cm]
0 & {\rm for} & i = j
\end{array}\right., \\[0.2cm]
M_{C}^{ij} &= \Bigg\{\begin{array}{cll}
\frac{\epsilon_i \cdot p_j}{\sigma_{ij}} & {\rm for} & i \neq j\\[0.2cm]
- \sum_{k \neq i} \frac{\epsilon_i \cdot p_k}{\sigma_{ik}}  & {\rm for}& i = j
\end{array}, \nonumber
\end{align}
with $\epsilon_i$ the polarization vector of the $i$-th gauge boson.
The reduced Pfaffian in Eq. \eqref{eq:IR_redpfaffian} is defined as the Pfaffian of the matrix 
whose entries are obtained from $M_{n}$ by removing the $k$-th row and $\ell$-th column and multiplied by 
$(-1)^{k+\ell}\sigma_{k\ell}^{-1}$.

Since the determinant of an antisymmetric matrix is the square of its Pfaffian, it is natural that graviton amplitudes
in Einstein-Hilbert gravity can also be written in the form \eqref{eq:general_amplitude_SE}, with the gauge theory 
factor $I_{L}$ shown in Eq. \eqref{eq:IL} replaced by a second copy of the reduced Pfaffian
\begin{align}
I_L &= {\rm Pf}' M_n,\nonumber \\[0.2cm]
I_R &= {\rm Pf}' M_n.
\end{align}
This exhibits the double-copy structure of graviton amplitudes, already 
found in many other physical setups \cite{Kawai:1985xq,Bern:2008qj,Bern:2010ue,Monteiro:2013rya,Naculich:2014naa,Vera:2014tda}.
Similarly, biadjoint scalar amplitudes can be obtained from Eq. \eqref{eq:general_amplitude_SE} substituting 
the reduced Pfaffian in $I_{R}$ by a second copy of the gauge theory factor \eqref{eq:IL}, thus implementing the 
zeroth copy prescription \cite{Monteiro:2014cda}.

\subsection{Fairlie's solution to the scattering equations}
\label{sec:fairlie_sln}

In this section we discuss a definite solution to the SE discussed by Fairlie in~\cite{Fairlie:2008dg} (see also \cite{Fairlie:1972zz}), 
which always exists for any multiplicity $n$. It has the form
\begin{align}
\sigma_j &= \frac{p_j^0 + p_j^3}{p_j^1 - i p_j^2}  =  \frac{\left(p_j^0 + p_j^3\right)\left(p_j^1 + i p_j^2\right)}{(p_j^1)^2 +  (p_j^2)^2 } 
\nonumber \\[0.2cm]
&=  \frac{\left(p_j^0 + p_j^3\right)\left(p_j^1 + i p_j^2\right)}{(p_j^0)^2 -  (p_j^3)^2 } = \frac{p_j^1 + i p_j^2}{p_j^0 - p_j^3},
\label{eq:sln_fairlie}
\end{align}
where we work in $D=4$ with the mostly-minus signature. 
Since $\sigma_{i}$ admits two expressions in terms of the momentum components, we can write two alternative 
identities to be satisfied by the differences $\sigma_{ij}$
\begin{align}
\sigma_{i j} \left(p_i^1 - i p_i^2\right) \left(p_j^0 - p_j^3\right)
&=  p_i \cdot p_j - p_i^0 p_j^3 +   p_i^3 p_j^0   - i p_i^1p_j^2 +  i p_i^2 p_j^1, \nonumber \\[0.2cm]
\sigma_{i j} \left(p_j^1 - i p_j^2\right) \left(p_i^0 - p_i^3\right) &=   - p_i \cdot p_j + p_j^0 p_i^3 -   p_j^3 p_i^0   + i p_j^1p_i^2 
-  i p_j^2 p_i^1.
\label{rep2}
\end{align}
Subtracting both equations, we arrive at the expression
\begin{align}
\left(p_i^1 - i p_i^2\right) \left(p_j^0 - p_j^3\right) - \left(p_j^1 - i p_j^2\right) \left(p_i^0 - p_i^3\right)
&= 2 \frac{p_i \cdot p_j}{\sigma_{i j} }.
\end{align}
We can use this identity to explicitly check that \eqref{eq:sln_fairlie} is indeed a solution to the SE. Summing over $j$ with $j \neq i$ we have
\begin{align}
  2 \sum_{\substack{j=1\\j\neq i }}^n \frac{p_i \cdot p_j}{\sigma_{i j} }&=\left(p_i^1 - i p_i^2\right) \sum_{\substack{j=1\\j\neq i }}^n 
\left(p_j^0 - p_j^3\right) - \left(p_i^0 - p_i^3\right) \sum_{\substack{j=1\\j\neq i }}^n \left(p_j^1 - i p_j^2\right) 
\nonumber\\[0.2cm]
&=- \left(p_i^1 - i p_i^2\right) \left(p_i^0 - p_i^3\right) 
 + \left(p_i^0 - p_i^3\right)  \left(p_i^1 - i p_i^2\right)=0,
\end{align}
where we have made use of momentum conservation.

It is possible to bring these solutions into a more physical representation if we use the following parametrization of on-shell momenta $p_j$ 
\begin{align}
p_j = p_{j}^{\perp} \left(\cosh{Y_j},\cos{\phi_j},\sin{\phi_j},\sinh{Y_j}\right),
\label{eq:ps_rapidity_azimuthal_def}
\end{align}
where $Y_j$ is the rapidity, $\phi_{j}$ the azimuthal angle, and the overall scale $p_{j}^{\perp}$ equals the modulus of the transverse 
component of the momentum.
To connect this representation with the one in terms of the $n$-punctured sphere, we notice that  $p_j$ can be alternatively written as
\begin{align}
p_j =  \omega_j \left(1,\mathbf{u}_{j}\right),
\label{eq:pj_uj}
\end{align}
where we have introduced the unit vector
\begin{align}
\mathbf{u}_{j} = (x_j,y_j,z_j), \hspace*{1cm} \mathbf{u}_{j}^{2}=1,
\end{align}
and $\omega_{j}$ is the energy of the $j$-th particle. Using this parametrization it is glaring how a null momentum is completely specified
by the energy of the particle and its direction of flight, corresponding to a point on 
$\mathbb{R} \times \mathbb{S}^2$. Points on the celestial sphere $\mathbb{S}^{2}$ can be parametrized either using stereographic coordinates $\zeta_j$ 
or the 
polar and azimuthal angles ($\theta_j,\phi_j)$. They are related by the following identities
\begin{align}
x_j&=\sin{\theta_j} \cos{\phi_j}=\frac{2 e^{Y_j} \cos{\phi_j}}{1+ e^{2 Y_j}} = \frac{\zeta_j + \bar{\zeta}_j}{1+ \zeta_j \bar{\zeta}_j},
\nonumber \\[0.2cm]
y_j&=\sin{\theta_j} \sin{\phi_j}= \frac{2 e^{Y_j} \sin{\phi_j}}{1+ e^{2 Y_j}} = i \frac{\bar{\zeta}_j-\zeta_j}{1+ \zeta_j \bar{\zeta}_j},\\[0.2cm]
z_j&=\cos{\theta_j}=\frac{e^{2Y_j}-1 }{1+ e^{2 Y_j}} = \frac{\zeta_j \bar{\zeta}_j -1}{1+ \zeta_j \bar{\zeta}_j},
\nonumber
\end{align}
which can be inverted to give
\begin{align}
\zeta_j &= e^{Y_j} e^{i \phi_j} ~=~ \frac{\sin{\theta_j}}{1 - \cos{\theta_j }} e^{i \phi_j}=\cot{\frac{\theta_j}{2}} e^{i \phi_j},
\nonumber \\[0.2cm]
\bar{\zeta}_j &= e^{Y_j} e^{-i \phi_j}=\frac{\sin{\theta_j}}{1 - \cos{\theta_j }} e^{-i \phi_j}=\cot{\frac{\theta_j}{2}} e^{- i \phi_j}.
\end{align}
This leads to the following parametrization of the particle momenta in terms of its energy and the stereographic coordinates on $\mathbb{S}^{2}$
\begin{align}
p_j= \omega_j \left(1, \frac{\zeta_j + \bar{\zeta}_j}{1+ \zeta_j \bar{\zeta}_j},i \frac{ \bar{\zeta}_j-\zeta_j}{1+ \zeta_j \bar{\zeta}_j},\frac{\zeta_j \bar{\zeta}_j-1}{1 + \zeta_j \bar{\zeta}_j}\right).
\end{align}

\begin{figure}
\vspace{-1.5cm}
\begin{center}
\includegraphics[scale=.8]{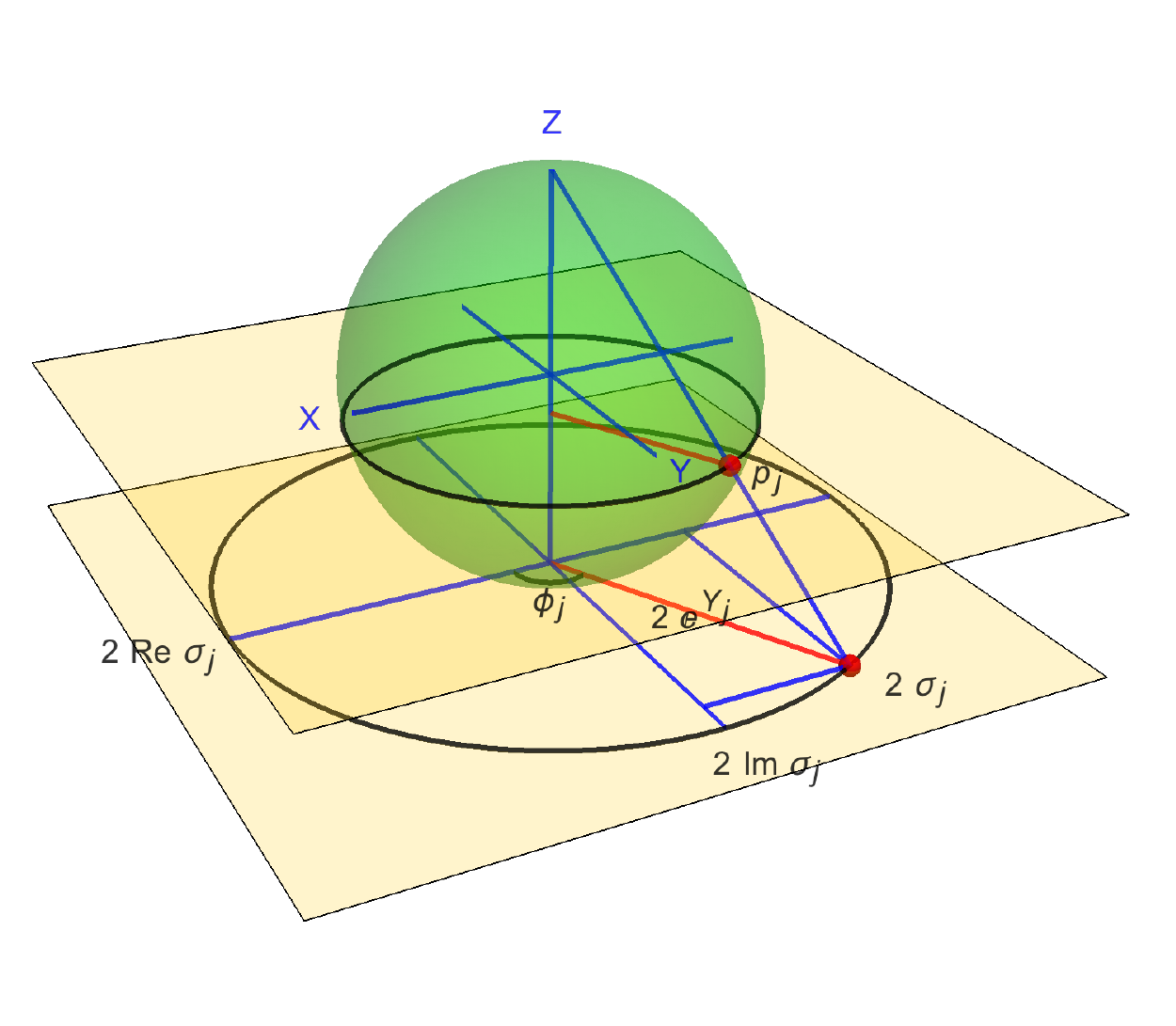}
\vspace{-1.2cm}
\caption{Geometric interpretation of the rapidity $Y_j$ and azimuthal angle $\phi_j$.}
\label{Stereo1}
\end{center}
\end{figure}

Using the previous representation of the particle momenta, we see that Fairlie's solution \eqref{eq:sln_fairlie} to the SE is simply given by  
\begin{align}
\sigma_j = \zeta_j = e^{Y_j + i \phi_j}.
\label{eq:punctures_position_Yphi}
\end{align}
Since we will make frequent use of this representation in the following, some remarks are in order. In Fig.~\ref{Stereo1} 
we have represented a point in the celestial sphere and its image on the complex plane whose origin coincides  with the south pole. 
The direction of flight of a particle with momentum $p_{j}$ labelled by the complex coordinate $\zeta_{j}$ is mapped onto the
point $2\sigma_{j}$ on that plane. At fixed rapidity $Y_{j}$, the points lie on a circumference of radius $2e^{Y_{j}}$ parametrized
by the azimuthal angle $\phi_{j}$. 

\section{Incoming momenta}

In this section we investigate the structure on the punctured sphere for the two incoming particles with momenta $p$ and $q$ in a general process 
in which the particles in the final state have momenta $p_{i}$ (with $i=1,\ldots,n-2$). We will consider the case when the two incoming particles' 
spatial momenta lie along the $z$ axis. It is convenient to work first with the parametrization in terms of rapidities and azimuthal angles introduced
in Eq. \eqref{eq:ps_rapidity_azimuthal_def}
\begin{align}
p&=\ell \left(\cosh{Y_p},\cos{\phi},\sin{\phi}, \sinh{Y_p}\right), \nonumber \\[0.2cm]
q&=\ell \left(\cosh{Y_q},-\cos{\phi},-\sin{\phi}, \sinh{Y_q}\right),
\label{eq:p,q_parametrization}
\end{align}
where we have set  both transverse momenta equal, $p^{\perp}=q^{\perp}\equiv \ell$. 
To study the limit of vanishing transverse momenta, we take $\ell\rightarrow 0$ and $|Y_{p}|,|Y_{q}|\rightarrow \infty$, 
while keeping the center-of-mass energy
\begin{align}
s=2p\cdot q = 2\ell^{2}\Big[1+\cosh{(Y_p-Y_q)}\Big]
\label{eq:def_s_rapidities}
\end{align}
finite.
This limit can be implemented by introducing a parameter $\epsilon$
\begin{align}
Y_{p}=-Y_{q}=-\log\epsilon,
\end{align}
that we eventually take to zero. A look at Eq. \eqref{eq:def_s_rapidities} shows that in order to keep $s$ finite we 
are forced to take the double scaling limit
\begin{align}
\epsilon\longrightarrow 0, \hspace*{0.5cm} \ell\longrightarrow 0 \hspace*{0.5cm} \mbox{with}\hspace*{0.5cm}
{\ell\over \epsilon}=\sqrt{s},
\label{eq:double_scaling}
\end{align} 
in which the incoming momenta take the form
\begin{align}
p&\longrightarrow  \frac{\sqrt{s}}{2} (1,0,0,1), \nonumber \\[0.2cm] 
q&\longrightarrow \frac{\sqrt{s}}{2} (1,0,0,-1).
\label{eq:incoming_momenta}
\end{align}

We can rephrase this double scaling in terms of the  position of the corresponding punctures on the sphere 
$\{\sigma_{p},\sigma_{q}\}$, which satisfy the identity 
\begin{align}
\frac{\sigma_p}{\sigma_q} + \frac{\sigma_q}{\sigma_p} = 2 - \frac{s}{\ell^2}.
\end{align}
Equation \eqref{eq:punctures_position_Yphi} shows that for small $\epsilon$ the two punctures are located on 
a small circle around the north and south poles of the Riemann sphere, which shrinks to a point when 
$\epsilon\rightarrow 0$, namely
\begin{align}
\sigma_p &= e^{Y_p + i \phi} = \frac{e^{i \phi}}{\epsilon} \longrightarrow \infty, \nonumber \\[0.2cm]
\sigma_q &= - e^{Y_q + i \phi}=  - \epsilon \, e^{i \phi} \longrightarrow 0.
\end{align}
These punctures can be alternatively labelled by the unit vectors $\mathbf{u}_{p}$ and $\mathbf{u}_{q}$ defined by 
Eq. \eqref{eq:pj_uj}. In our case, they take the form
\begin{align}
\mathbf{u}_{p} &= \left(\frac{\cos{\phi}}{\cosh{Y_p}},\frac{\sin{\phi}}{\cosh{Y_p}}, \tanh{Y_p}\right),\nonumber \\[0.2cm]
\mathbf{u}_{q} &= \left(-\frac{\cos{\phi}}{\cosh{Y_q}},-\frac{\sin{\phi}}{\cosh{Y_q}}, \tanh{Y_q}\right),
\end{align}
whose projections onto the equatorial plane lie on circles with respective radii
\begin{align}
R_{p} &= \frac{1}{\cosh{Y_p}}, \nonumber \\[0.2cm] R_{q} &= \frac{1}{\cosh{Y_q}},
\end{align}
which shrink to zero as $|Y_{p,q}|\rightarrow \infty$ (i.e., $\epsilon\rightarrow 0$).

\begin{figure}
\vspace{-1.cm}
\begin{center}
\includegraphics[scale=.7]{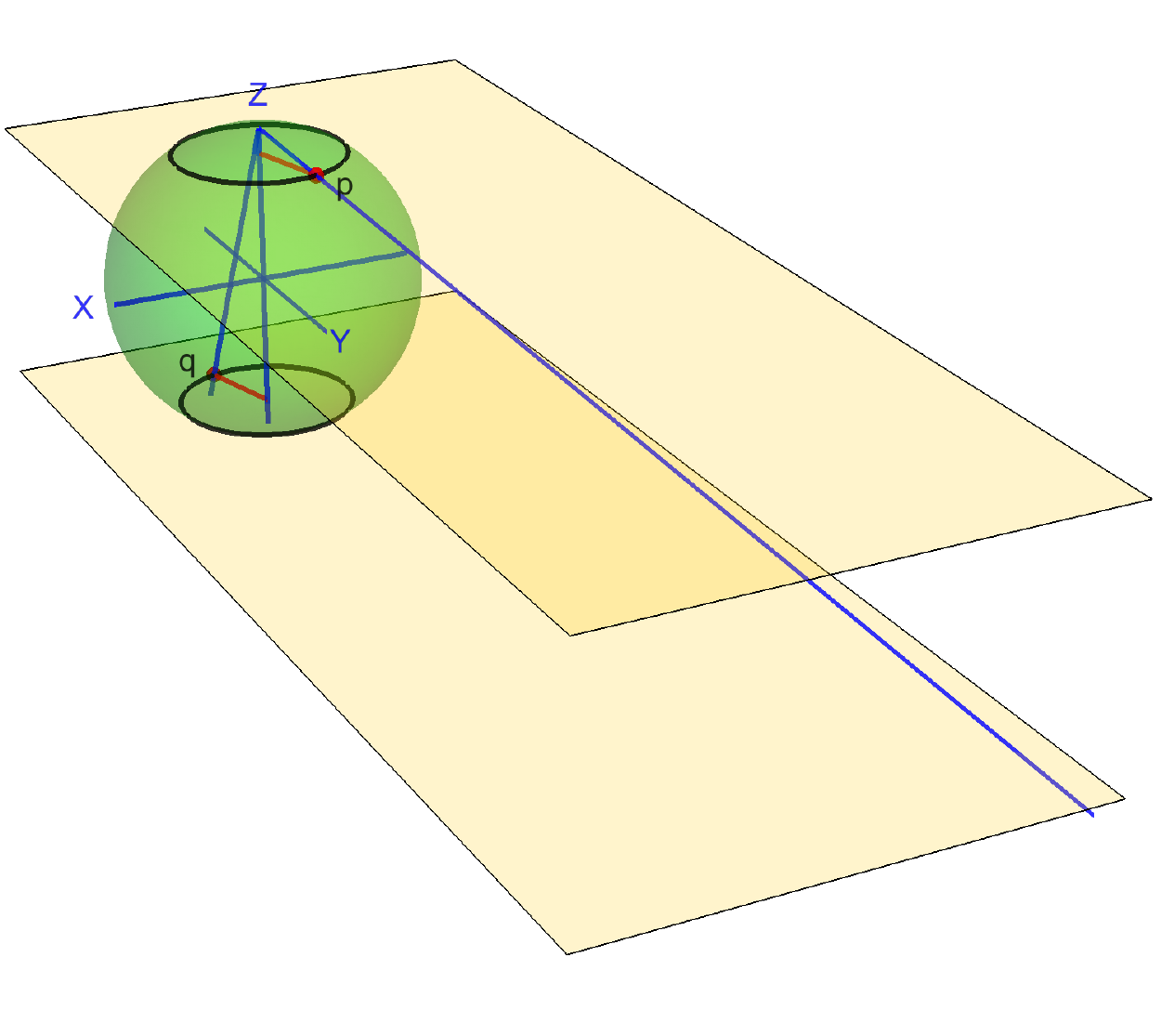}
\vspace{-1.cm}
\caption{Stereographic projection for two incoming particles whose momenta lie close to the $z$ axis.}
\label{Stereo2}
\end{center}
\end{figure}
The geometric setup for the configuration discussed here is illustrated in Fig.~\ref{Stereo2}, where we show
the punctures associated with the incoming particles very close to the north and south poles of the Riemann
sphere. The value of $\phi$ 
is ambiguous for points on the $z$ axis and without loss of generality we can set it to zero from now on, 
since this angle is a mere artefact of the way we take the limit. On the complex plane this
means that the limits $\sigma_q\rightarrow 0$ and $\sigma_p\rightarrow \infty$ are taken 
along the real axis.

\section{Sudakov representation of the scattering equations: the four-point case}

After introducing our setup and conventions, we turn to study the formulation of the SE formalism in terms of Sudakov
parameters. We begin with the simplest case, that of a general four-point scattering amplitude with 
incoming and outgoing momenta respectively given by $p$, $q$ and $p'$, $q'$, which are constrained by momentum conservation
\begin{align}
p + q - p' - q' =0.
\label{eq:momentum_conservation4}
\end{align}
We parametrize the two incoming momenta $p$ and $q$  as explained in the previous Section. 

\subsection{Punctures on the Riemann sphere}

In the CHY formalism \cite{Cachazo:2013hca,Cachazo:2013iea}, the momenta
$\{p, q, p', q'\}$ are mapped into the moduli space of spheres with four punctures, 
located respectively at the points $\{\sigma_p, \sigma_q,\sigma_{p'} ,\sigma_{q'}\} \in \mathbb{C}\mathbb{P}^1$. This 
is implemented by the identities 
\begin{align}
p^\mu  &= \oint\limits_{|z-\sigma_p|=\epsilon}\, \frac{dz}{2 \pi i} 
\omega^\mu (z), \nonumber \\[0.2cm]
q^\mu  &= \oint\limits_{|z-\sigma_q|=\epsilon}\, \frac{dz}{2 \pi i} 
\omega^\mu (z), \nonumber\\ 
{p'}^\mu  &= - \hspace*{-0.5cm}\oint\limits_{|z-\sigma_{p'}|=\epsilon} \frac{dz}{2 \pi i}\, 
\omega^\mu (z), \label{Localizations} \\[0.2cm]  
{q'}^\mu  &= - \hspace*{-0.5cm}\oint\limits_{|z-\sigma_{q'}|=\epsilon} \frac{dz}{2 \pi i}\,
\omega^\mu (z), \nonumber
\end{align}
where the meromorphic function $\omega^{\mu}(z)$ is fully determined by the condition that it has poles at the location of the punctures
whose residues are the corresponding particle momenta
\begin{align}
\omega^\mu (z) &= \frac{p^\mu}{z-\sigma_p}+\frac{q^\mu}{z-\sigma_q}-\frac{{p'}^\mu}{ z-\sigma_{p'}}-\frac{{q'}^\mu}{ z-\sigma_{q'}}.
\end{align}

The incoming momenta are parametrized as shown in Eq. \eqref{eq:p,q_parametrization} with $\phi=0$.
For the outgoing particles, on the other hand, we write their momenta introducing a Sudakov~\cite{Sudakov:1954sw} representation.
Due to momentum conservation \eqref{eq:momentum_conservation4}, it is enough to parametrize the combination
\begin{align}
q_1 &\equiv p - p' = \alpha \,  p + \beta \, q + \mathbf{q}_1, 
\end{align}
with
\begin{align}
\mathbf{q}_1 = q_1^\perp \left(0,\cos{\theta_1},\sin{\theta_1},0\right).
\end{align} 
Then, the momentum $p'$ can be written as
\begin{align}
p' = p - q_1  &= \ell\bigg(\left(1-\alpha\right) \cosh{Y_p} - \beta \cosh{Y_q},0,0,\left(1-\alpha\right) \sinh{Y_p} - \beta \sinh{Y_q}\bigg) \nonumber \\
&+ \bigg(0, \left(1-\alpha+ \beta\right) \ell - q_1^\perp \cos{\theta_{1}}, - q_1^\perp \sin{\theta}_1,0\bigg),
\label{eq:p1_4p}\\[0.2cm]
\longrightarrow \left({\sqrt{s}\over 2}\right.&\left.(1-\alpha-\beta),-q_{1}^{\perp}\cos\theta_{1},-q_{1}^{\perp}\sin\theta_{1},
{\sqrt{s}\over 2}(1-\alpha+\beta)\right),
\nonumber
\end{align}
where in the last expression we have taken the double scaling limit \eqref{eq:double_scaling}.
From this we read the particle energy
\begin{align}
\omega_{p'}=\frac{\sqrt{s}}{2} \left(1-\alpha-\beta\right),
\end{align}
whereas the on-shell condition leads to 
\begin{align}
0=p'^{2}&=- s (1-\alpha) \beta - (q_{1}^{\perp})^{2} \hspace*{1cm} 
\Longrightarrow \hspace*{1cm} |Q_{1}|^{2}\equiv (q_{1}^{\perp})^{2}=s(\alpha-1)\beta,
\label{eq:on-shell_Q1^2}
\end{align}
where we have introduced the notation 
\begin{align}
Q_{j}=q_j^\perp e^{i \theta_j}.
\label{eq:Q_jDef}
\end{align}

We repeat the same calculation for the momentum $q'$ of the second outgoing particle. In terms of 
the Sudakov parameters, it reads
\begin{align}
q' = q + q_1  &= \ell\bigg(\alpha \cosh{Y_p} + \left(1+\beta\right) \cosh{Y_q},0,0,\alpha \sinh{Y_p} + \left(1+\beta\right) \sinh{Y_q}\bigg) \nonumber \\
&+ \bigg(0, \left(\alpha -1- \beta\right) \ell  + q_1^\perp \cos{\theta}_1,
q_1^\perp \sin{\theta}_1,0\bigg) \\[0.2cm]
\longrightarrow \left({\sqrt{s}\over 2}\right.&\left.(1+\alpha+\beta)
,q_{1}^{\perp}\cos\theta_{1},q_{1}^{\perp}\sin\theta_{1},{\sqrt{s}\over 2}(-1+\alpha-\beta)
\right),
\nonumber
\end{align}
where we have reabsorbed a sign in a shift of $\theta_{1}$ by $\pi$. Comparing with the expression for $p'$ in Eq. \eqref{eq:p1_4p}
we see that this reflects the fact that, in the center-of-mass frame, the two outgoing particles fly in opposite directions
and therefore their azimuthal angles differ by $\pi$. 
The energy of the particle is given by 
\begin{align}
\omega_{q'} =  \frac{\sqrt{s}}{2} \left(1+\alpha+\beta\right),
\end{align}
whereas the on-shell condition $q'^{2}=0$ 
leads to the constraint
\begin{align}
0=q'^{2}&=s \alpha(1+\beta) - (q_{1}^{\perp})^{2} \hspace*{1cm} 
\Longrightarrow \hspace*{1cm} |Q_{1}|^{2}\equiv (q_{1}^{\perp})^{2}=s\alpha(1+\beta).
\label{eq:on-shell_Q1^2_q'}
\end{align} 
Consistency with the value of $|Q_{1}|^{2}$ found from the on-shell condition $p'^{2}=0$ in Eq. 
\eqref{eq:on-shell_Q1^2} implies that $\alpha$ 
and $\beta$ are not independent, but rather satisfy 
\begin{align}
\alpha+\beta=0.
\label{eq:a+b=0}
\end{align}
This condition implies that
\begin{align}
\omega_{p'}=\omega_{q'}={\sqrt{s}\over 2},
\end{align}
as it behoves a four particle scattering in the center-of-mass frame.  

Let us recall that for the four-point function, the SE only have one solution. Thus, it is enough to consider 
Fairlie's solution \eqref{eq:sln_fairlie} reviewed in Section \ref{sec:fairlie_sln}. 
This being the case, the complex coordinate of the puncture in the sphere associated with the momentum $p'$ is given by
\begin{align}
\sigma_{p'} &\equiv e^{Y_{p'} + i \phi_{p'}}
=\frac{ Q_1}{ \beta \sqrt{s}}= \sqrt{\frac{1-\alpha}{\alpha}} e^{i(\theta_1+\pi)},
\label{eq:sigma_p'4part}
\end{align}
where in using \eqref{eq:on-shell_Q1^2} to write the result in terms of $Q_{1}$ we have made a choice of phase
for the square root. 
In addition, the projection of the associated unit vector $\mathbf{u}_{p'}$
\begin{align}
\mathbf{u}_{p'} &={2\over \sqrt{s}}\Big(q_{1}^{\perp}\cos(\theta_{1}+\pi),q_{1}^{\perp}\sin(\theta_{1}+\pi),
{\sqrt{s}\over 2}(1-2\alpha)\Big),
\end{align}
onto the equatorial plane lies on a circumference with radius
\begin{align}
R_{p'}={2\sqrt{\alpha(1-\alpha)}},
\label{eq:Rp'_4p}
\end{align}
where we have used the on-shell condition \eqref{eq:on-shell_Q1^2}.
Going to the Riemann sphere representation, the complex coordinate of the puncture associated with the particle of momentum
$q'$ is
\begin{align}
\sigma_{q'} &\equiv 
 e^{Y_{q'}+ i \phi_{q'}}=  
\frac{ Q_1}{(1-\alpha) \sqrt{s}} = \sqrt{\frac{\alpha}{1-\alpha}} e^{i\theta_1},
\end{align}
where our choice of phase is consistent with the one used for $\sigma_{p'}$ in Eq. \eqref{eq:sigma_p'4part}. 
Thus, we conclude
\begin{align}
\sigma_{q'}=-{1\over \sigma_{p'}^{*}}=\sqrt{\alpha\over 1-\alpha}e^{i\theta_{1}},
\label{eq:solution_4p}
\end{align}
indicating that the two punctures are located on antipodal points on the sphere. 
This becomes obvious when computing the components of the unit vector $\mathbf{u}_{q'}$
\begin{align}
\mathbf{u}_{q'} &= {2\over \sqrt{s}} \left( q_{1}^\perp \cos\theta_1,
q_1^\perp \sin\theta_1,-{\sqrt{s}\over 2} (1-2\alpha)\right).
\end{align}
Now, since after imposing \eqref{eq:a+b=0} we see that $\mathbf{u}_{q'}=-\mathbf{u}_{p'}$, the projection of both 
vectors on the equatorial plane defines the same loci, namely a circumference with radius 
[cf. Eq. \eqref{eq:Rp'_4p}]
\begin{align}
R_{p'}=R_{q'}=2\sqrt{\alpha(1-\alpha)},
\end{align}
whereas their components along the direction of the incoming particles are
\begin{align}
Z_{p'}=-Z_{q'}=1-2\alpha.
\end{align}

\begin{figure}
\vspace{-2.4cm}
\begin{center}
\includegraphics[scale=.8]{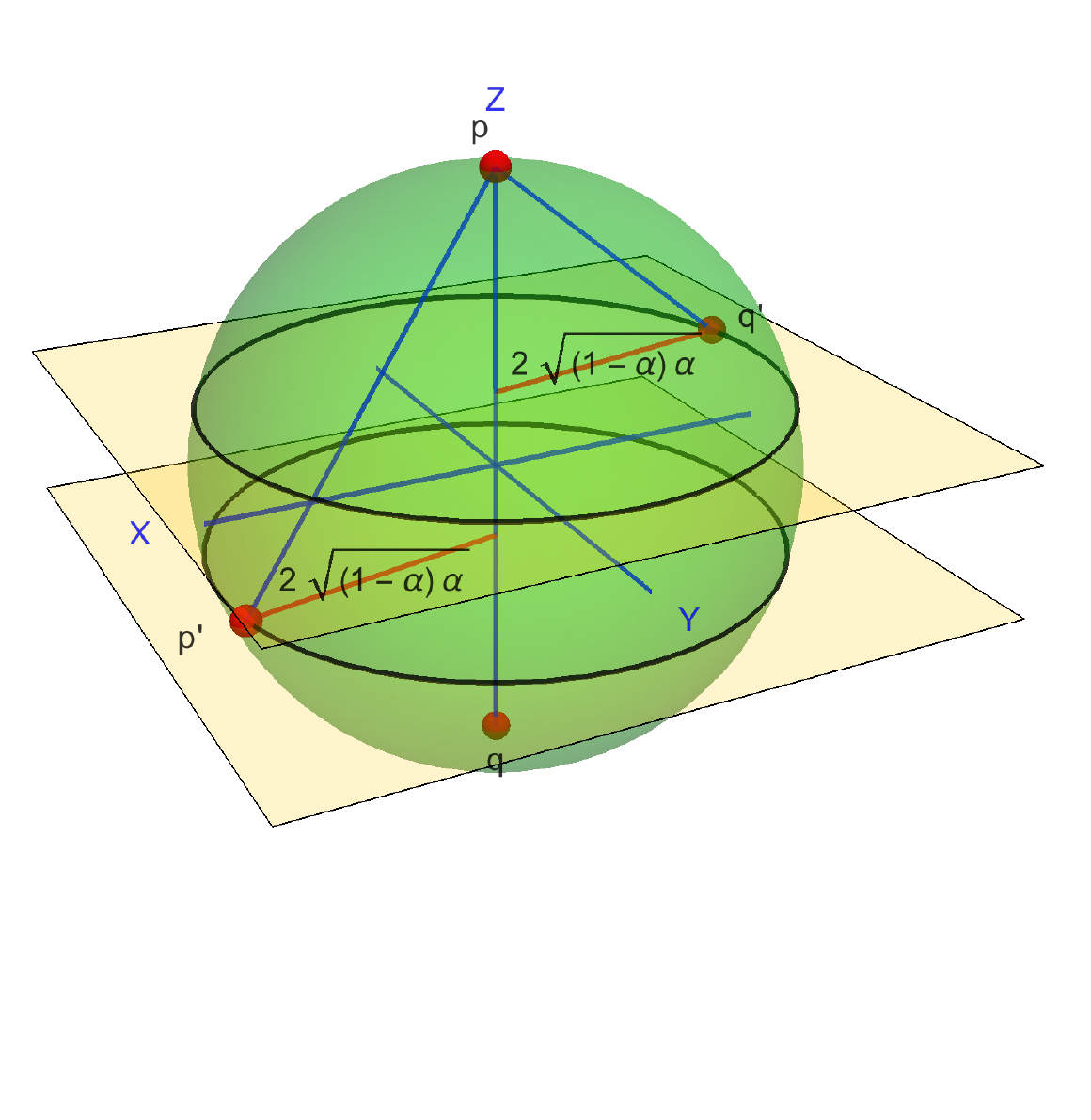}
\vspace{-2.4cm}
\caption{Punctures on the Riemann sphere for the four particle scattering with momenta $p + q \longrightarrow p' + q'$.
In the limits $\alpha\rightarrow 0,1$ the outgoing punctures collide with the incoming ones located at the poles.}
\label{Stereo3}
\end{center}
\end{figure}
In Fig.~\ref{Stereo3} we provide a pictorial example of the parametrization proposed above. The boundary of the 
moduli space of the sphere with four punctures is approached in the limits $\alpha\rightarrow 1$ or
$\alpha\rightarrow 0$. 
They correspond to the coincidence limit in which the punctures associated with the outgoing particles collide with
those of the incoming ones, located at the north and south pole of the Riemann sphere. In the case $\alpha={1\over 2}$ 
the radii $R_{p'}=R_{q'}=2\sqrt{\alpha(1-\alpha)}$ reach the maximum value and the outgoing particles are emitted along the equatorial plane. 

\subsection{The scattering equations and the four-point amplitude}

In order to write the SE using the Sudakov parametrization, we need to compute the Mandelstam invariants
\eqref{eq:mandelstam_inv}
where, according to our conventions $p_{1}=p$, $p_{2}=q$, $p_{3}=-p'$, and $p_{4}=-q'$. 
For four particle scattering, they have the following explicit form
\begin{align}
s_{pq}&=s_{p' q'}= s,    \\
s_{pp'}&=s_{qq'}=- q_1^2=- t=s \alpha, \\
s_{pq'} &=s_{qp'}=-u=s (1- \alpha).
\end{align}
Since $s$ is the only dimensionful quantity available, we use rescaled variables
\begin{align}
s_{ij}&=s\widehat{s}_{ij}, \nonumber \\[0.2cm]
Q_{i}&= \sqrt{s}\widehat{Q}_{i}. 
\label{eq:rescaled_qty}
\end{align}
It is straightforward to check that the SE associated with $p$ is trivially satisfied
\begin{align}
\frac{{\cal S}_{p}}{s}&={\widehat{s}_{pq}\over \sigma_{pq}}-{\widehat{s}_{pp'}\over \sigma_{pp'}}-
{\widehat{s}_{pq'}\over \sigma_{pq'}}=0,
\end{align}
since we have $\sigma_p = \infty$. In the case of the SE associated to $q$
\begin{align}
\frac{{\cal S}_{q}}{s} &=  \frac{\widehat{s}_{pq}}{\sigma_{qp}}- \frac{\widehat{s}_{qq'}}{\sigma_{qq'}} - \frac{\widehat{s}_{q p'}}{\sigma_{qp'}}, 
\end{align}
we have a nontrivial cancellation. The first term vanishes again because $\sigma_p=\infty$ and 
we can use the explicit expressions
\begin{align}
\sigma_{p'}&=-{\widehat{Q}_{1}\over \alpha}, \nonumber \\[0.2cm]
\sigma_{q'}&={\widehat{Q}_{1}\over 1-\alpha},
\label{eq:sigmas_p'q'}
\end{align}
together with $\sigma_{q}=0$. 
Using this Sudakov representation, it is easy to check that the SE is fulfilled
\begin{align}
\frac{{\cal S}_{q}}{s}=\frac{\widehat{s}_{qq'}}{\sigma_{q'}}+\frac{\widehat{s}_{q p'}}{\sigma_{p'}}=
-\frac{\alpha (1-\alpha)}{\widehat{Q}_1}+\frac{(1-\alpha)\alpha}{\widehat{Q}_1}=0,
\end{align}
and similarly for the two remaining SE
\begin{align}
\frac{{\cal S}_{p'}}{s} &=  \frac{\widehat{s}_{p' q}}{\sigma_{p'}}- \frac{\widehat{s}_{p'q'}}{\sigma_{p'q'}} 
=-{\alpha(1-\alpha)\over \widehat{Q}_{1}}+{\alpha(1-\alpha)\over \widehat{Q}_{1}}=0, 
\nonumber \\[0.2cm]
\frac{{\cal S}_{q'}}{s}&=\frac{\widehat{s}_{qq'}}{\sigma_{q'}} - \frac{\widehat{s}_{p'q'}}{\sigma_{q'p'}} 
=-{\alpha(1-\alpha)\over \widehat{Q}_{1}}+{\alpha(1-\alpha)\over \widehat{Q}_{1}}=0.
\end{align}

The Sudakov representation provides a very convenient framework for the evaluation of scattering amplitudes in the CHY formalism, 
notably simplifying the computations. To illustrate this, we focus now on the calculation of
the four-point amplitude in a $\varphi^{3}$ scalar theory. According to the general prescription given in 
\cite{Cachazo:2013hca,Cachazo:2013iea}, the amplitude can be written as the following integral supported on the solution
to the SE
\begin{align}
\mathcal{A}_4^{\varphi^3} &=  \int d z_{p'} \delta \left({\cal S}_{p'}\right) 
\frac{z_{pq}^2 z_{qq'}^2 z_{q'p}^2}{\left(z_{pq} z_{qq'} z_{q'p'} z_{p'p} \right)^2} 
\nonumber  \\[0.2cm]
&=\int\frac{d z_{p'}}{(z_{p'}-\sigma_{q'})^2} \delta \left({s_{p'q}\over z_{p'}}-{s_{p'q'}\over z_{p'}-\sigma_{q'}}\right),
\label{eq:4p_amplitude_CHY}
\end{align}
where all gauge generators are taken to be equal to one.
Here we have partially fixed the SL(2,$\mathbb{C}$) invariance by setting $z_{p} = \infty$ and $z_{q} =0$ while 
leaving the third one 
\begin{align}
z_{q'}=\sigma_{q'}={Q_{1}\over (1-\alpha)\sqrt{s}},
\end{align}
free. The integral defining the scattering amplitude has just one integration left
over the position of the $p'$ puncture. To carry out this integral, we notice that the argument of the delta function
has a single root located at [see \eqref{eq:sigmas_p'q'}]
\begin{align}
z_{p'}=-{Q_{1}\over \alpha\sqrt{s}}.
\label{eq:zp'_solution_4p}
\end{align}
Evaluating the derivative of $S_{p'}$ with respect to the integration variable at the zero \eqref{eq:zp'_solution_4p}
gives
\begin{align}
\mathcal{J}\equiv \left.\frac{\partial S_{p'}}{\partial z_{p'}}\right|_{z_{p'}=-{Q_{1}\over \alpha\sqrt{s}}}
= \frac{s^{2} \alpha^{2}(\alpha-1)}{Q_{1}^2} + \frac{s\alpha^{2}(1-\alpha)^{2}}{Q_{1}^{2}}
= \frac{s^2 \alpha^3  (\alpha-1)}{Q_1^2},
\end{align}
so we can simply write
 \begin{align}
\mathcal{A}_4^{\varphi^3} &=  \int  
d z_{p'}\left[z_{p'}-{Q_{1}\over (1-\alpha)\sqrt{s}}\right]^{-2} \frac{Q_1^2}{s^2 \alpha^3  (\alpha-1)}
\delta \left(z_{p'} + \frac{Q_1}{\alpha \sqrt{s}}\right)
 \nonumber\\[0.2cm]
&= \left[\frac{s \alpha^2 (1-\alpha)^2}{Q_1^2}\right] \left[\frac{Q_1^2}{s^2 \alpha^3  (\alpha-1)}\right] = 
  \frac{ (\alpha-1) }{s \alpha}  = \frac{1}{s} +\frac{1}{t}.
 \end{align} 
Notice that the phase introduced in $Q_1$, which contains the azimuthal angle dependence,  cancels out in the final expression for the amplitude. This is only natural, since $\theta_{1}$ can be set to zero by using the residual SL(2,$\mathbb{C}$) transformations
leaving invariant the position of the punctures associated with the incoming particles. 
Using this Sudakov parametrization, we see how the boundary of the 4-punctured sphere corresponding to the limit 
$\alpha \rightarrow 0$ is dominated by the $t=0$ pole, while at the other branch of the boundary $\alpha\rightarrow 1$ the amplitude vanishes. At the equator $\alpha={1\over 2}$ the amplitude is completely dominated by the pole at $s=0$.

\section{Sudakov representation of the five-point amplitude}

After the analysis of the four-point amplitude, we turn to the scattering of five particles which enjoys some more interesting 
features, mainly the existence of a second solution to the SE besides Fairlie's. To fix notation, we will now study a 
generic five-point scattering amplitude of particles with momenta $p + q \rightarrow p' + k + q'$ satisfying the momentum conservation
identity
\begin{align}
p+q-p'-k-q'=0.
\end{align}

\subsection{Location of the punctures}

The mapping between particle momenta and the puncture positions is provided by the relations listed in 
Eq.~\eqref{Localizations} supplemented with the one for $k$
\begin{align}
k^\mu  &= - \oint_{\left|z-\sigma_k \right|=\epsilon} \frac{dz}{2 \pi i} 
\omega^\mu (z),
\end{align}
where now the meromorphic function $\omega^{\mu}(z)$ is given by
\begin{align}
\omega^\mu (z) &= \frac{p^\mu}{z-\sigma_p}+\frac{q^\mu}{z-\sigma_q}-\frac{{p'}^\mu}{ z-\sigma_{p'}}
-\frac{k^\mu}{ z-\sigma_{k}}-\frac{{q'}^\mu}{ z-\sigma_{q'}}.
\end{align}
To parametrize the momenta, we introduce two pairs of Sudakov parameters $\{\alpha_{1},\beta_{1}\}$ and $\{\alpha_{2},\beta_{2}\}$
such that
\begin{align}
q_1 &= p - p' = \alpha_1 p + \beta_1 q + \mathbf{q}_{1}, \nonumber \\[0.2cm]
q_2 &= q' - q = \alpha_2 p + \beta_2 q + \mathbf{q}_{2},\\[0.2cm]
k &= q_1 - q_2 = \left(\alpha_1 - \alpha_2\right) p + \left(\beta_1 - \beta_2\right) q + 
\mathbf{q}_1 - \mathbf{q}_{2}, \nonumber
\end{align}
where the transverse vectors have components
\begin{align}
\mathbf{q}_i&=q_i^\perp \Big(0,\cos{\theta_i},\sin{\theta_i},0\Big).
\end{align}
Using again the notation introduced in Eq. \eqref{eq:Q_jDef}, and taking the double scaling limit  
\eqref{eq:double_scaling}, we have
\begin{align}
p'= p - q_1  &= \ell\Big((1-\alpha_1) \cosh{Y_p} - \beta_1 \cosh{Y_q},0,0,\left(1-\alpha_1\right) \sinh{Y_p} - \beta_1 \sinh{Y_q}\bigg) \nonumber \\[0.2cm]
&+ \Big(0,(1-\alpha_1+ \beta_{1})\ell- q_{1}^{\perp}\cos{\theta_{1}},-q_{1}^{\perp} \sin{\theta_1},0\Big) \\[0.2cm]
\longrightarrow \left({\sqrt{s}\over 2}\right.&\left.(1-\alpha_{1}-\beta_{1}),-q_{1}^{\perp}\cos\theta_{1},-q_{1}^{\perp}\sin\theta_{1},
{\sqrt{s}\over 2}(1-\alpha_{1}+\beta_{1})\right).
\nonumber
\end{align}
A similar analysis can be repeated for the remaining two outgoing particles. In terms of the Sudakov parameters, 
their momenta take the form
\begin{align}
q'=q + q_2&=\ell\Big(\alpha_2 \cosh{Y_p} + (1+\beta_2) \cosh{Y_q},0,0,\alpha_2 \sinh{Y_p} + (1+\beta_2) \sinh{Y_q}\Big)  
\nonumber\\[0.2cm]
&+ \Big(0, (\alpha_2-\beta_2-1)\ell+q_{2}^{\perp} \cos{\theta_{2}},q_{2}^{\perp} \sin{\theta_{2}},0\Big) 
\nonumber \\[0.2cm]
\longrightarrow \left({\sqrt{s}\over 2}\right.&\left.(1+\alpha_{2}+\beta_{2}),q_{2}^{\perp}\cos\theta_{2},q_{2}^{\perp}\sin\theta_{2},
{\sqrt{2}\over 2}(-1+\alpha_{2}-\beta_{2})\right), \\[0.2cm]
k = q_1 - q_2 
&= \ell\bigg(\left(\alpha_1-\alpha_2\right) \cosh{Y_p} + \left(\beta_1-\beta_2\right) \cosh{Y_q},0,0,\left(\alpha_1-\alpha_2\right) \sinh{Y_p} + \left(\beta_1-\beta_2\right) \sinh{Y_q}\bigg) \nonumber \\
&+ \bigg(0,  q_1^\perp \cos{\theta}_1 - q_2^\perp \cos{\theta}_2,q_1^\perp \sin{\theta}_1 - q_2^\perp \sin{\theta}_2,0\bigg) 
\nonumber \\[0.2cm]
\longrightarrow \Big(
{\sqrt{s}\over 2}&(\alpha_{1}+\beta_{1}-\alpha_{2}-\beta_{2}),q_{1}^{\perp}\cos\theta_{1}-q_{2}^{\perp}\cos\theta_{2},
q_{1}^{\perp}-q_{2}^{\perp}\sin\theta_{2},{\sqrt{s}\over 2}(\alpha_{1}-\beta_{1}-\alpha_{2}+\beta_{2})\Big).
\nonumber
\end{align}

The associated energies are read off these expressions to be 
\begin{align}
\omega_{p'} &= \frac{\sqrt{s}}{2} \left(1-\alpha_1-\beta_1\right), \nonumber \\[0.2cm]
\omega_{q'} &= {\sqrt{s}\over 2}(1+\alpha_{2}+\beta_{2}), \\[0.2cm]
\omega_{k} &= {\sqrt{s}\over 2}(\alpha_{1}+\beta_{1}-\alpha_{2}-\beta_{2}), \nonumber
\end{align}
which obviously satisfy energy conservation, $\omega_{p'}+\omega_{q'}+\omega_{k}=\sqrt{s}$. In addition, the
on-shell condition for the outgoing momenta fixes the magnitude of the transverse momenta in terms of the 
Sudakov parameters
\begin{align}
p'^{2}&=0 \hspace*{1cm} \Longrightarrow \hspace*{1cm} |Q_{1}|^{2}=s(\alpha_{1}-1)\beta_{1}, \nonumber \\[0.2cm]
q'^{2}&=0 \hspace*{1cm} \Longrightarrow \hspace*{1cm} |Q_{2}|^{2}=s\alpha_{2}(1+\beta_{2}), 
\label{eq:Qs5p}\\[0.2cm]
k^{2}&=0  \hspace*{1cm} \Longrightarrow \hspace*{1cm} |Q_{1}-Q_{2}|^{2}=s(\alpha_{1}-\alpha_{2})(\beta_{1}-\beta_{2}).
\nonumber
\end{align}
In fact, combining them we find a further identity
\begin{align}
Q_{1}Q_{2}^{*}+Q_{1}^{*}Q_{2}= s(\alpha_{2}-\beta_{1}+\alpha_{1}\beta_{2}+\alpha_{2}\beta_{1}).
\end{align}
It is important to stress at this point that, unlike the situation encountered in the four-point amplitude, here
the on-shell conditions for the outgoing particles do not lead to consistency identities restricting the values
of the Sudakov parameters. Thus, whereas in the case of four particles the identity \eqref{eq:a+b=0} implies 
the existence of a single independent Sudakov parameter, in the five-point amplitude the four parameters  
remain independent.

The coordinates of the punctures associated with each momenta corresponding to Fairlie's solution are given by
\begin{align}
\sigma_{p'} &=
\frac{ Q_1}{ \beta_1 \sqrt{s}}=\sqrt{\frac{\alpha_1 - 1}{\beta_1}} e^{i \theta_1} 
=e^{Y_{p'} + i \phi_{p'}},
\label{PhysicalSolutionp}
\\[0.2cm]
\sigma_{q'} &=  
\frac{ Q_2}{\left(1+ \beta_2\right) \sqrt{s}}=\sqrt{\frac{\alpha_2}{1+\beta_2}} e^{i \theta_2}=e^{Y_{q'}+ i \phi_{q'}},
\label{PhysicalSolutionq} \\[0.2cm]
\sigma_{k} &=\frac{ Q_1- Q_2 }{\left(\beta_1- \beta_2 \right) \sqrt{s}} =
\frac{ \sqrt{(\alpha_1-1)\beta_1} e^{i \theta_1}- \sqrt{(1+\beta_2) \alpha_2} e^{i \theta_2} }{\beta_1- \beta_2 } = e^{Y_k + i \phi_k}, \label{PhysicalSolutionk}
\end{align}
which are the stereographic coordinates labelling the directions of flight of the particles. In order to visualize the position
of these punctures, it is convenient to use the unity vectors
\begin{align}
\mathbf{u}_{p'}&= {2\over \sqrt{s}(1-\alpha_{1}-\beta_{1})}\Big(- q_{1}^{\perp} \cos{\theta_{1}},
-q_{1}^{\perp} \sin{\theta_{1}},{\sqrt{s}\over 2}(1-\alpha_{1}+\beta_{1})\Big),
\nonumber \\[0.2cm]
\mathbf{u}_{q'}&= {2\over \sqrt{s}(1+\alpha_{2}+\beta_{2})}\Big(q_2^\perp \cos{\theta_{2}},
q_2^\perp \sin{\theta_{2}}, {\sqrt{s}\over 2} (-1+\alpha_{2}-\beta_{2})\Big), 
\label{eq:vectors_punctures_a5p}\\[0.2cm]
\mathbf{u}_{k} &= {2\over \sqrt{s}(\alpha_{1}+\beta_{1}-\alpha_{2}-\beta_{2})}\Big(q_1^\perp \cos{\theta_{1}} - q_2^\perp \cos{\theta_{2}}, q_1^\perp \sin{\theta_{1}} 
- q_2^\perp \sin{\theta_{2}},{\sqrt{s}\over 2}(\alpha_{1}-\beta_{1}-\alpha_{2}+\beta_{2})\Big). \nonumber
\end{align}
Using the expression for $q_{i}^{\perp}$ given in Eq. \eqref{eq:Qs5p}, we see that the projections of $\mathbf{u}_{p'}$ and
$\mathbf{u}_{q'}$ lie onto the equatorial plane on circumferences with radii
\begin{align}
R_{p'}&=2{\sqrt{(\alpha_{1}-1)\beta_{1} \over (1-\alpha_{1}-\beta_{1})^{2}}}, \nonumber \\[0.2cm]
R_{q'}&=2{\sqrt{\alpha_{2}(1+\beta_{2}) \over (1+\alpha_{2}+\beta_{2})^{2}}}. 
\end{align}
For the momentum $k$, we just need to notice that since $\theta_{1}$ and $\theta_{2}$ are respectively 
the arguments of $Q_{1}$ and $Q_{2}$
\begin{align}
q_{1}^{\perp}\cos\theta_{1}-q_{2}^{\perp}\cos\theta_{2}&={\rm Re\,}\Big(Q_{1}-Q_{2}\Big), \nonumber \\[0.2cm]
q_{1}^{\perp}\sin\theta_{1}-q_{2}^{\perp}\sin\theta_{2}&={\rm Im\,}\Big(Q_{1}-Q_{2}\Big).
\end{align}
Hence, the equatorial projection of $\mathbf{u}_{k}$ lies on a circumference of radius
\begin{align}
R_{k}&={2|Q_{1}-Q_{2}|\over \alpha_{1}+\beta_{1}-\alpha_{2}-\beta_{2}}=2\sqrt{(\alpha_{1}-\alpha_{2})(\beta_{1}-\beta_{2})
\over (\alpha_{1}+\beta_{1}-\alpha_{2}-\beta_{2})^{2}}.
\end{align}

\begin{figure}[t]
\vspace{-2.5cm}
\begin{center}
\includegraphics[scale=.6]{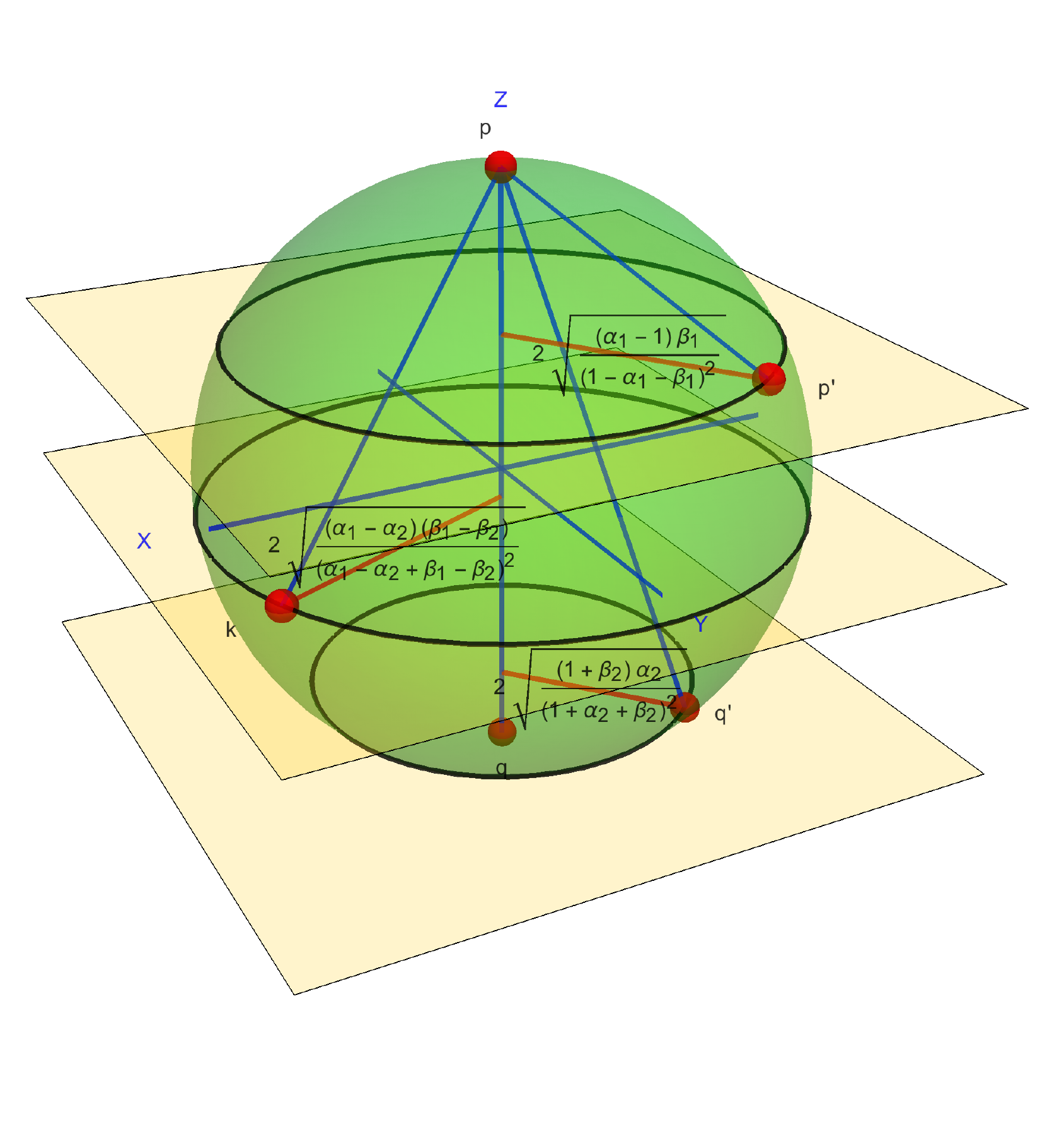}
\vspace{-1.5cm}
\caption{Punctures on the Riemann sphere for the five-particle amplitude.}
\label{Stereo4}
\end{center}
\end{figure}
In Fig.~\ref{Stereo4} we show a typical configuration for the five punctures on the Riemann sphere.  
Several factorization channels can be identified in the expressions given in this Section. An interesting one corresponds to 
$\beta_1, \alpha_2 \rightarrow 0$, with both $\alpha_1$ and $-\beta_2$ not close to $1$. This limit 
sends the puncture associated with $p'$ to the north pole, while the puncture for 
$q'$ approaches the south pole. In this limit the puncture for $k$ remains at the equator 
whenever $\alpha_{1}+\beta_{2}=0$. Alternatively, we can keep $\sigma_{k}$ at the equator by taking 
$\alpha_{1},-\beta_{2}\rightarrow 1$, with $\beta_1$ and $\alpha_2$ not close to $0$. On the other hand, 
the puncture associated with $k$ moves to the south pole in the limit $\alpha_1\rightarrow \alpha_2$ and to the north pole 
if $\beta_{1} \rightarrow \beta_{2}$.

\subsection{Scattering equations}

To write the SE, we begin by computing the Mandelstam invariants \eqref{eq:mandelstam_inv} in terms of the Sudakov parameters
for the five-point amplitude
\begin{align}
s_{pq} &= s, \hspace*{2.5cm}  
s_{p'k} = - s (\alpha_2+\beta_2), \hspace*{1.5cm}
s_{q' k} = s  \left(\alpha_1+\beta_1\right), \nonumber \\[0.2cm]
s_{pp'} &= - s \beta_1, \hspace*{1.8cm} s_{qq'} = s \alpha_2, \hspace*{3.15cm}s_{pk} = s (\beta_1 - \beta_2),
\nonumber \\[0.2cm]
s_{qk} &= s (\alpha_1 - \alpha_2), \hspace*{0.8cm} s_{pq'} =  s (1+\beta_2), \hspace*{2.1cm}
s_{qp'} = s (1-\alpha_1), \\[0.2cm]
s_{p' q'} &= s (1-\alpha_1 + \alpha_2 -\beta_1 +\beta_2).
\nonumber
\end{align}
By inverting these relations, it is possible to express the Sudakov parameters in terms of the invariants as
\begin{align}
 s \, \alpha_1 &= s_{q'k} + s_{p p'}, \hspace{1cm}
s \, \alpha_2 = s_{qq'},\\[0.2cm]
s \, \beta_1 &= - s_{pp'},\hspace{1.8cm}
s \, \beta_2 = - s_{p' k} - s_{q q'}.
\end{align}

We know that for the five-point amplitude there must be two different solutions. One of them is the one found by Fairlie 
\cite{Fairlie:2008dg,Fairlie:1972zz} that we have expressed in Eq. \eqref{PhysicalSolutionk} in terms of Sudakov parameters. 
To find the second one, we write the ansatz
\begin{align}
\sigma_{p'} &= C_p\,\widehat{Q}_{1}, \nonumber \\[0.2cm]
\sigma_{q'} &= C_q \,\widehat{Q}_{2},
\label{eq:rescalings_Cpq}
\end{align}
with $C_{p}$ and $C_{q}$ two complex constants and we use the rescaled quantities defined in Eq. \eqref{eq:rescaled_qty}.
A first condition comes from complying with the SE associated to $q$,
\begin{align}
{\cal S}_q &\equiv \frac{1-\alpha_1 }{\sigma_{p'}}+\frac{\alpha_2}{\sigma_{q'}}+ \frac{\alpha_1 - \alpha_2}{\sigma_{k}} =0,
\end{align}
which determines $\sigma_{k}$ to be
\begin{align}
\sigma_k
&=  (\alpha_2 -\alpha_1)\left(\frac{1-\alpha_1}{C_p {\widehat Q}_1} + \frac{\alpha_2}{C_q {\widehat Q}_2}\right)^{-1}.
\end{align}
Now we impose the SE associated to $q'$, which reads
\begin{align}
{\cal S}_{q'} &\equiv \frac{\alpha_2}{\sigma_{q'}}+ \frac{1-\alpha_1 + \alpha_2 -\beta_1 +\beta_2}{\sigma_{p'q'}}
+ \frac{\alpha_1+\beta_1}{\sigma_{kq'}}=0,
\end{align}
leading to the relation
\begin{align}
\frac{\sigma_{k}}{\sigma_{q'}} = \frac{(\alpha_2-\alpha_1)\sigma_{p'}}{\alpha_2 \sigma_{p'}+(1-\alpha_1)\sigma_{q'}} = 
\frac{(\alpha_2-\alpha_1-\beta_1) \sigma_{p'} +(1+\beta_2) \sigma_{q'}}{\alpha_2 \sigma_{p'}+(1-\alpha_1 -\beta_1 +\beta_2) \sigma_{q'}}.
\label{sigmak5point}
\end{align}
Using the on-shell relations \eqref{eq:Qs5p}, this equation can be equivalently written as
\begin{align}
\alpha_2 \beta_1 \sigma_{p'}^2 - \left({\widehat Q}_1  {\widehat Q}_2^* + {\widehat Q}_1^*  {\widehat Q}_2\right) \sigma_{p'} \sigma_{q'} + \frac{|{\widehat Q}_1|^2 |{\widehat Q}_2 |^2}{\alpha_2 \beta_1} \sigma_{q'}^2 = 0.
\label{Quadratic5point}
\end{align}
Assuming $\sigma_{q'}\neq 0$, this is a quadratic equation for the ratio ${\sigma_{p'}\over \sigma_{q'}}$ whose coefficients are 
expressed only in terms of the Sudakov parameters. Its two solutions are given by
\begin{align}
\frac{\sigma_{p'}^{(\pm)}}{\sigma_{q'}^{(\pm)}} = 
{1\over 2 \alpha_2 \beta_1}\left[{\widehat Q}_1  {\widehat Q}_2^* + {\widehat Q}_1^*  {\widehat Q}_2
\pm \sqrt{\left({\widehat Q}_1  {\widehat Q}_2^* + {\widehat Q}_1^*  {\widehat Q}_2\right)^2 - 4 |{\widehat Q}_1|^2 |{\widehat Q}_2|^2}
\right], 
\end{align}
which admits the simpler form
\begin{align}
\frac{\sigma_{p'}^{(+)}}{\sigma_{q'}^{(+)}} = \frac{{\widehat Q}_1  {\widehat Q}_2^*   }{\alpha_2 \beta_1}, \nonumber \\[0.2cm]
\frac{\sigma_{p'}^{(-)}}{\sigma_{q'}^{(-)}} = \frac{ {\widehat Q}_1^*  {\widehat Q}_2}{\alpha_2 \beta_1} .
\end{align}
Being solutions to a quadratic equation with real coefficients, they are complex conjugate of each other. 
Using now the second equation in \eqref{eq:rescalings_Cpq}, together with \eqref{sigmak5point} and the on-shell conditions
\eqref{eq:Qs5p}, we arrive at the following
expression of the solution $\sigma_{i}^{(+)}$ to the SE  
\begin{align}
\sigma_{p'}^{(+)}&= C_q \, \frac{(1+\beta_2)}{\beta_1} \widehat{Q}_1, \nonumber \\[0.2cm]
\sigma_{q'}^{(+)}&= C_q \, \widehat{Q}_2, \\[0.2cm]
\sigma_{k}^{(+)} &= C_q \, \frac{(1+\beta_2)}{\beta_1 - \beta_2} \left(\widehat{Q}_1 - \widehat{Q}_2\right).
\nonumber
\end{align}

To fix the undetermined constant $C_{q}$ we identify $\sigma_{i}^{(+)}$ with Fairlie's solution \eqref{PhysicalSolutionk}. This
fixes $C_{q}$ to be  
\begin{align}
C_{q} &= {e^{-i\theta_{2}}\over 1+\beta_2}.
\label{eq:Cq_5p}
\end{align}
In order to understand the presence of the phase in this expression, we should 
point out that, in setting $\sigma_{p}=\infty$ and $\sigma_{q}=0$, we only partially fixed the SL(2,$\mathbb{C}$)
invariance of the moduli space of punctured spheres. This leaves us with complex rescalings as the residual invariance. We can 
make use of this freedom to set the phase of the constant $C_{q}$ as in 
\eqref{eq:Cq_5p}, which geometrically corresponds to a change in the origin of the azimuthal angles in the Riemann sphere. 
Our choice, which sets $\sigma_{q'}^{(\pm)}$ on the real axis, 
leads to a more symmetric form of the two solutions to the SE for the five-point amplitude
\begin{align}
\sigma_{p'}^{(+)} &=  {\sigma_{p'}^{(-)*}} =\frac{{\widehat Q}_1 e^{-i \theta_2}}{\beta_1} 
=\sqrt{\frac{\alpha_1-1}{\beta_1}}  e^{i (\theta_1-\theta_2 + \pi)}, \nonumber \\[0.2cm]
\sigma_{q'}^{(+)} &= {\sigma_{q'}^{(-)*}}= \frac{{\widehat Q}_2 e^{-i \theta_2}}{1+\beta_2} 
= \sqrt{\frac{\alpha_2}{1+\beta_2}}, \label{eq:gen_sol_A5_sigmas} \\[0.2cm]
\sigma_{k}^{(+)} &= {\sigma_{k}^{(-)*}}= \frac{({\widehat Q}_1 - {\widehat Q}_2) e^{-i \theta_2}}{\beta_1 - \beta_2} ~=~ \frac{\sqrt{(\alpha_1-1)\beta_1} e^{i (\theta_1-\theta_2)}- \sqrt{\alpha_2 (1+\beta_2)} }{\beta_1-\beta_2}.
\nonumber
\end{align}

The localization of the punctures on the Riemann sphere can be also given in terms of the unit vectors
\begin{align}
\mathbf{u}_{p'}^{(\pm)} &= {1\over 1-\alpha_1- \beta_1}
\Big( -2 \sqrt{(\alpha_1-1)\beta_1}  \cos\gamma,  \mp 2 \sqrt{(\alpha_1-1)\beta_1}  \sin\gamma,  1-\alpha_1+\beta_1 \Big),
\label{pS2}\\
\mathbf{u}_{q'}^{(\pm)} &= {1\over \alpha_2+ \beta_2 +1}\Big( 2 \sqrt{\alpha_2 (1+\beta_2)},  0,\alpha_2-\beta_2 -1 \Big),
\label{qS2}\\
\mathbf{u}_{k}^{(\pm)}  &= {1\over \alpha_1+\beta_1-\alpha_2- \beta_2 } 
\Big( 2 \sqrt{(\alpha_1-1)\beta_1}  \cos\gamma -  \sqrt{\alpha_2 (1+\beta_2)},  \nonumber \\[0.2cm]
&\hspace*{3.8cm}\mp 2 \sqrt{(\alpha_1-1)\beta_1}  \sin\gamma,\alpha_1-\beta_1-\alpha_2+\beta_2 \Big),
\label{kS2}
\end{align}
where we have defined $\gamma=\theta_{1}-\theta_{2}$. As announced, $\sigma^{(+)}_{i}$ corresponds to 
Fairlie's solution, after choosing the origin 
of azimuthal angles such that $\theta_{2}=0$ in Eq. \eqref{eq:vectors_punctures_a5p}. The second solution $\sigma^{(-)}_{i}$
is obtained by reflecting the first one with respect to the $y=0$ plane.

\subsection{Scalar scattering amplitude}

Having obtained the two solutions to the SE, we are now ready to calculate the five-point amplitude 
for the $\varphi^{3}$ scalar theory. Using the same partial fixing of SL(2,$\mathbb{C}$) as in the calculation of the four-point
amplitude in Eq. \eqref{eq:4p_amplitude_CHY}, we are left with the computation of the integral over the position of the
punctures associated with $p'$ and $q'$, namely
\begin{align}
\mathcal{A}_{5}^{\varphi^3} &=  \int d z_{p'} d z_{q'}\,\delta\left({\cal S}_{p'}\right) 
\delta\left({\cal S}_{q'}\right) 
\frac{z_{pq}^2 z_{qk}^2 z_{kp}^2}{\left(z_{pq} z_{qq'} z_{q'k} z_{kp'} z_{p'p} \right)^2}\nonumber \\[0.2cm]
&= \int d z_{p'} d z_{q'}\,\delta\left({\cal S}_{p'}\right) \delta\left({\cal S}_{q'}\right) 
\frac{ z_{k}^2 }{z_{q'}^2 z_{q'k}^2 z_{kp'}^2 }.
\end{align}
To solve the delta function, we have to calculate the Jacobian 
\begin{align}
{\cal J} &= \frac{\partial {\cal S}_{p'} }{\partial \sigma_{p'}} \frac{\partial {\cal S}_{q'} }{\partial \sigma_{q'}} 
- \frac{\partial {\cal S}_{p'} }{\partial \sigma_{q'}} \frac{\partial {\cal S}_{q'} }{\partial \sigma_{p'}}.
\end{align}
Things can be made simpler if we rewrite the SE associated to $q'$ in the form
\begin{align}
{1\over s}\mathcal{S}_{q'} &= \frac{\alpha_2}{\sigma_{q'}}+ \frac{(1-\alpha_1 + \alpha_2 -\beta_1 +\beta_2) }{\sigma_{p'q'}}
+ \frac{\left(\alpha_1+\beta_1\right)}{\sigma_{kq'}}  \\
&= \frac{\left( \alpha_2 \sigma_{p'}   + (1-\alpha_1  -\beta_1 +\beta_2) \sigma_{q'} \right)}{ \sigma_{p'q'} \sigma_{kq'}} 
\left[\frac{\sigma_{k}}{\sigma_{q'}}
- \frac{\left( \alpha_2 -\alpha_1-\beta_1\right) \sigma_{p'}   
+ (1+\beta_2) \sigma_{q'} }{ \alpha_2 \sigma_{p'}   + (1-\alpha_1  -\beta_1 +\beta_2) \sigma_{q'} }\right].
\nonumber
\end{align}
What makes this expression useful is that 
we have isolated the zero due to Eq.~(\ref{sigmak5point}). 
We can then write one of the derivatives on support on the SE as 
\begin{align}
\left.{1\over s}{\partial \mathcal{S}_{q'} \over \partial \sigma_{q'}}\right|_{\rm SE}  
&= {\alpha_2 \sigma_{p'}   + (1-\alpha_1  -\beta_1 +\beta_2) \sigma_{q'}\over  \sigma_{p'q'} \sigma_{kq'}}
 \frac{\partial  }{\partial \sigma_{q'}}  \left[{\sigma_{k} \over \sigma_{q'}}
- {\left( \alpha_2 -\alpha_1-\beta_1\right) \sigma_{p'}   
+ (1+\beta_2) \sigma_{q'} \over \alpha_2 \sigma_{p'}   + (1-\alpha_1  -\beta_1 +\beta_2) \sigma_{q'}}\right] 
\nonumber\\[0.2cm]
&= {(\alpha_1-\alpha_2-\beta_1) \alpha_2 \sigma_{p'} +\big[ (\alpha_1-\alpha_2) \beta_2 + (\alpha_2-1) \beta_1 \big]\sigma_{q'} 
\over \alpha_2 \sigma_{p'}+(1-\alpha_1)\sigma_{q'}} \left(
{\sigma_{p'}\over \sigma_{q'} \sigma_{p'q'} \sigma_{kq'}} \right),
\end{align}
where we have used Eqs.~(\ref{sigmak5point}) and (\ref{Quadratic5point}). This can be further simplified 
by reintroducing $\sigma_k$ to write
\begin{align}
\left.{\partial \mathcal{S}_{q'} \over \partial \sigma_{q'}}
\right|_{\rm SE}  &=  s{ (1+\beta_2) \sigma_{q'}^2 +\alpha_2 \sigma_k \sigma_{p'} \over \sigma_{q'}^2 \sigma_{p'q'} \sigma_{q' k}}.
\label{eq:dsdsigmaq'}
\end{align}

We now repeat the same procedure for the SE associated to $p'$, isolating the contribution to the zero
\begin{align}
{1\over s}\mathcal{S}_{p'} &= 
{(\alpha_2 -\beta_1 +\beta_2)  \sigma_{p'} + (1-\alpha_1) \sigma_{q'} \over \sigma_{p'q'} \sigma_{p'k}}
 \left[ \frac{\sigma_{k}}{\sigma_{p'}}
  + \frac{(\alpha_1 -1 -\alpha_2 - \beta_2)  \sigma_{q'}  + \beta_1   \sigma_{p'}}{  ( \alpha_2 -\beta_1 +\beta_2)  \sigma_{p'} + (1-\alpha_1) \sigma_{q'} }\right],
\end{align}
and differentiating with respect to $\sigma_{p'}$, we find
\begin{align}
\left.{\partial \mathcal{S}_{p'}\over \partial \sigma_{p'}}\right|_{\rm SE}
&=s{\beta_{1}\sigma_{p'}^{2}+(\alpha_{1} -1)\sigma_{k}\sigma_{q'} 
\over \sigma_{p'}^2 \sigma_{p'q'} \sigma_{p'k}}.  
\label{eq:dsdsigmap'}
\end{align}
Note that the two partial derivatives \eqref{eq:dsdsigmaq'} and \eqref{eq:dsdsigmap'}
can be mapped to each other by the replacements
\begin{align}
p'&\longleftrightarrow q', \nonumber \\[0.2cm]
-\beta_{1} &\longleftrightarrow 1+ \beta_{2}, \label{ppqprelation} \\[0.2cm]
\alpha_{1} &\longleftrightarrow 1-\alpha_{2}. 
\nonumber
\end{align} 
Similar techniques allow to obtain the remaining two derivatives
\begin{align}
\left.{\partial \mathcal{S}_{q'} \over \partial \sigma_{p'}}\right|_{\rm SE}
&=   s{  \left( \alpha_1 + \beta_1\right) (1-\alpha_1 +\alpha_2  -\beta_1 +\beta_2)  \over (\beta_1 -\beta_2)
\sigma_{k} +(\alpha_1-\alpha_2) 
\sigma_{p'} }\left( {\sigma_{k} \over \sigma_{p'q'} \sigma_{q'k} }\right), \nonumber \\[0.2cm]
\left.{\partial \mathcal{S}_{p'} \over \partial \sigma_{q'}} \right|_{\rm SE}
  &=   s{ (1-\alpha_1 + \alpha_2 - \beta_1 + \beta_2) ( \alpha_2 +\beta_2)   
    \over  (\beta_1 -\beta_2)\sigma_{k}  +(\alpha_1-\alpha_2)\sigma_{q'}}
    \left( {\sigma_{k}\over \sigma_{p'q'} \sigma_{p'k}  }\right),
\end{align}
which are also related by the transformations (\ref{ppqprelation}).

We can return to the calculation of the amplitude, which can be written as the sum over the two solutions complex conjugate of each other:
\begin{align}
\mathcal{A}_5^{\varphi^3}
&= \int d z_{p'} d z_{q'} \, {\cal J}^{-1}\delta\big(z_{p'}-\sigma_{p'}\big) \delta\big(z_{q'}-\sigma_{q'}\big) 
\frac{ z_{k}^2 }{z_{q'}^2 z_{q'k}^2 z_{kp'}^2 } + {\rm c.c.} \nonumber \\[0.2cm]
&={2\over s^{2}}\mbox{Re}\left[\left({ \sigma_{p'q'}^2 \over \sigma_{q'k} \sigma_{p' k}}\right) {1 \over \mathcal{L}-\mathcal{R}}\right]
\label{eq:A5_final1}\\[0.2cm]
&=  {1\over s^{2}}\left[\frac{1}{\alpha_1+\beta_1}  - \frac{1}{\alpha_2+\beta_2}
+\frac{1}{(\alpha_1+\beta_1) \beta_1} - \frac{1}{\beta_1 \alpha_2}
+\frac{1}{\alpha_2 (\alpha_2+\beta_2) }\right].
\nonumber 
\end{align}
Here we have used the notation
\begin{align}
{\cal L} &\equiv  \left[\beta_1 \frac{\sigma_{p'}}{\sigma_{k}} 
  +       (\alpha_1 -1)  \frac{ \sigma_{q'} }{\sigma_{p'}} \right]
  \left[\alpha_2+(1+\beta_2) \frac{\sigma_{q'}^2}{\sigma_{k}\sigma_{p'}}\right], \\[0.2cm]
{\cal R} &\equiv {  (1-\alpha_1 +\alpha_2  -\beta_1 +\beta_2)^2 
  \left( \alpha_1 + \beta_1\right) ( \alpha_2 +\beta_2)   
  \sigma^2_{q'}  \over \big[(\beta_1 -\beta_2)\sigma_k +(\alpha_1-\alpha_2) \sigma_{p'} \big] \big[( \beta_1 -\beta_2)\sigma_k  +(\alpha_1-\alpha_2) \sigma_{q'} \big]},
\end{align}
as well as the explicit expressions for the positions of the punctures given in Eq. \eqref{eq:gen_sol_A5_sigmas}.  
In fact, the on-shell conditions \eqref{eq:Qs5p} can be written in the form
\begin{align}
2\cos(\theta_{1}-\theta_{2})=  \frac{ \alpha_2- \beta_1+  \alpha_1 \beta_2 +  \alpha_2 \beta_1}{ \sqrt{( \alpha_1-1)  \beta_1} \sqrt{(1+\beta_2)  \alpha_2}},
\end{align}
which is useful for the numerical evaluation of our expressions and 
shows how the relative phase depends on the Sudakov variables. 

In fact, it is possible to write an alternative expression for the amplitude \eqref{eq:A5_final1} as
\begin{align}
\mathcal{A}_{5}^{\varphi^3}
={2\over s^2} \mbox{Re\,} \left[\left({ \sigma_{p'} \over \sigma_{q'}} \right){ 1 \over L \widetilde{L} 
-R\widetilde{R}}\right],
 \label{last}
\end{align} 
where
\begin{align}
L &= {\sigma_{p'k}\over \sigma_{p'q'}} \left[(\alpha_1 - 1) \frac{\sigma_{q'}}{\sigma_{p'}} + \beta_1 \frac{\sigma_{p'}}{\sigma_{q'}}\right],\nonumber \\[0.2cm]
R &= \left({\sigma_{p'}\sigma_{p'k}\over \sigma_{p'q'}}\right)
{(1-\alpha_1+\alpha_2-\beta_1+\beta_2) (\alpha_1+\beta_1)\over 
(\alpha_1 - \alpha_2 + \beta_1)\sigma_{p'} - (1 + \beta_2) \sigma_{q'}},
\end{align}
and the quantities with tilde are defined by implementing the replacements  
\eqref{ppqprelation} in the form 
\begin{align}
\widetilde{\mathscr{O}}\Big(\alpha_1,\alpha_2,\beta_1,\beta_2,\theta_{1}-\theta_{2}\Big) =  
\mathscr{O}\Big(1-\alpha_2, 1-\alpha_1,-1-\beta_2,-1-\beta_1,\theta_{2}-\theta_{1}\Big).
\end{align}
The reason behind the existence of such a simple representation is the freedom to redefine the phase in the projective variables. This is part of the residual SL(2,${\mathbb C}$) freedom present in our approach, after
fixing the punctures associated with the two incoming particles. One interesting feature of Eq. \eqref{last} 
is related to the mathematical properties of the zeros of this amplitude, an issue which has
already been explored in \cite{Jimenez:2016sgc,Jimenez:2017kqu}. This will be further 
investigated in a future publication. 

\section{Conclusions and outlook}

We have presented a first complete analysis on the use of Sudakov variables in the context of the CHY calculation of scattering amplitudes. These amplitudes are represented as integrals with support on the solution to the scattering equations. Using these variables and a particular frame for the two incoming particles, it is possible to clearly identify the solutions to the scattering equations as punctures on the Riemann sphere parametrized by the rapidity and azimuthal angle, defined on the transverse plane to the collision axis of the incoming particles, of each on-shell particle. The punctures for the emitted particles are then living on circles parametrized by one Sudakov variable, in the four-point case, and four Sudakov variables for five-particles amplitude. In this formulation of the CHY calculation the final expression for a scalar amplitude has a very simple structure, given in terms of the position of the
punctures which are complex numbers carrying a phase defined as the difference of azimuthal angles of the emitted particles. 

In a future work we will generalize the Sudakov representation for a $n$-point amplitude and 
show the different multi-particle factorization limits which are naturally parametrized in this approach. 
The connection among gravitational and Yang-Mills amplitudes in this approach from the point of view of Regge kinematics~\cite{SabioVera:2011wy,Vera:2012ds,Johansson:2013nsa,Johansson:2013aca} is also of interest, together with the corresponding soft theorems~\cite{Vera:2014tda,Schwab:2014xua,He:2014bga,Bianchi:2014gla,Cachazo:2015ksa,DiVecchia:2015bfa}. 
Besides, it would be interesting to interpret the role of the gluing operator recently 
investigated in~\cite{Roehrig:2017gbt} in terms of Sudakov variables. Certainly, the relevance of this operator for the calculation of higher-loop amplitudes is still to be investigated and exploring kinematical limits, such as multi-Regge kinematics where the Sudakov representation is most useful, could be a possible route to understand its meaning.

\section*{Acknowledgements}

We would like to thank David Skinner, Piotr Tourkine and  Ellis Y. Yuan for very useful discussions.  A.S.V. would like to thank David Skinner for the warm hospitality at the Department of Applied Mathematics and Theoretical Physics (DAMTP), at the University of Cambridge, during the summer of 2017.  G.C., A.S.V. and D.M.J. acknowledge support from the Spanish Government grants FPA2015-65480-P, FPA2016-78022-P and Spanish MINECO Centro de Excelencia Severo Ochoa Programme (SEV-2016-0597).  
M.A.V.-M. has been partially supported by Spanish Government grant FPA2015-64041-C2-2-P and gratefully acknowledges 
the hospitality of the KEK Theory Center during the completion of this work.

\end{document}